\documentclass[journal]{IEEEtran}

\usepackage[dvipsnames]{xcolor}
\usepackage{epsfig,epsf,epstopdf}
\graphicspath{ {./Figs/} }
\usepackage[cmex10]{amsmath}
\usepackage{amsfonts}
\usepackage{amssymb}
\usepackage{algorithm}
\usepackage[noend]{algpseudocode}
\makeatletter
\def\BState{\State\hskip-\ALG@thistlm}
\makeatother
\usepackage{cite}
\usepackage{paralist}
\usepackage{color}
\usepackage{mathrsfs}
\usepackage{multirow}
\usepackage{subfigure}
\usepackage{graphicx}
\usepackage{caption}
\usepackage[font=small,justification=justified,singlelinecheck=false]{caption}

\definecolor{blue}{rgb}{0, 0.1, 0.8}

\ifCLASSINFOpdf
 
\else
 
\fi

\begin{document}

\title{Resilient Containment Control of Heterogeneous Multi-Agent Systems Against Unbounded Sensor and Actuator Attacks}

\author{Shan Zuo,
\IEEEmembership{Member,~IEEE},
        Yi Zhang, and
        Yichao Wang
\thanks{}
\thanks{Shan Zuo, Yi Zhang and Yichao Wang are with the Department of Electrical and Computer Engineering, University of Connecticut, CT 06269, USA; (E-mails: shan.zuo@uconn.edu; yi.2.zhang@uconn.edu; yichao.wang@uconn.edu).}
}

\maketitle

\begin{abstract}
Accurate local state measurement is important to ensure the reliable operation of distributed multi-agent systems (MAS). Existing fault-tolerant control strategies generally assume the sensor faults to be bounded and uncorrelated. In this paper, we study the ramifications of allowing the sensor attack injections to be unbounded and correlated. These malicious sensor attacks may bypass the conventional attack-detection methods and compromise the cooperative performance and even stability of the distributed networked MAS. Moreover, the attackers may gain access to the actuation computing channels and manipulate the control input commands. To this end, we consider the resilient containment control problem of general linear heterogeneous MAS in the face of correlated and unbounded sensor attacks, as well as general unbounded actuator attacks. We propose an attack-resilient control framework to guarantee the uniform ultimate boundedness of the closed-loop dynamical systems and preserve the bounded containment performance. Compared with existing literature addressing bounded faults and/or disturbances that are unintentionally caused in the sensor and actuator channels, the proposed control protocols are resilient against unknown unbounded attack signals simultaneously injected into sensor and actuator channels, and hence are more practical in the real-world security applications. A numerical example illustrates the efficacy of the proposed result, by highlighting the resilience improvement over the conventional cooperative control method.
\end{abstract}

\begin{IEEEkeywords}
Attack-resilient control, correlated attacks, heterogeneous MAS, unbounded attacks.        
\end{IEEEkeywords}

\section{Introduction}
Multi-agent systems (MAS) increasingly rely on distributed control paradigms with information exchanged among local controllers of agents on a sparse communication network \cite{1,2,3}. Such distributed cooperative control manner has broad applications in various engineering problems \cite{4,5,6} due to its high flexibility, adaptability, and scalability. The sparse communication networks among agents, however, pose security vulnerabilities as local agents lack the global perspective with limited information gathered from their neighbors. Malicious attackers could simultaneously launch attacks, at times in a coordinated fashion, on the sensors and actuators to undermine the cooperative performance and even stability of the dynamical systems \cite{7,8,9,10,11,12}. It is, therefore, necessary to seek reliable and secure remedies against malicious attacks.

The first approach to address attacks on MAS is to first detect the compromised agents \cite{13,14,15}. One could then either recover or remove the compromised agents. This approach generally has strict requirements on the number of compromised agents and/or the graphical connectivity of the communication network. Alternatively, distributed resilient method is recently studied to preserve an acceptable level of consensus performance by mitigating the propagated impact of external faults, noises, and/or disturbances \cite{16,17,18,19,20,21,22,23,24,25,26,27,28,29,30}. The main idea is to devise locally distributed control approaches to enhance the self-resilience of the MAS under malicious attacks, instead of detecting and recovering/removing the corrupted agents. Existing distributed resilient control methods for MAS mostly deal with bounded disturbances, noises, and/or faults. This may not be practical since the real-world attack injections could be purposely launched to undermine their targets to the maximum extent and, hence, cannot be assumed bounded \cite{15}. To summarize, existing solutions to address attacks have either strict limitations on the number of compromised agents, or connectivity of the communication topology, or do not address both sensor and actuator attacks in a unified manner, or assume bounded attack injections.

Accurate state measurement or estimation is important to guarantee the reliable operation of large-scale cyber-physical systems. Take the power grid for example, state estimation is needed to estimate the power grid state by analyzing meter measurements and power system models. The conventional attack-detection methods usually work well for independent and non-interacting bad measurements. It is shown in \cite{8} that intelligent attackers may exploit the power system configuration to launch simultaneous and coordinated attacks, which introduce correlated errors into state variables, and hence successfully bypass existing bad measurement detection techniques. The vulnerability assessment and consequences of power system state estimation with respect to such unobservable or undetectable false data injection attacks are presented in \cite{11,12}. For large-scale cyber-physical systems that are used in critical infrastructure, distributed attack-resilient control protocols are needed to address maliciously interacting sensor attack injections. Actuator faults are usually inevitable in realistic complicated systems and may deteriorate the system performance. Fault-tolerant control has been widely investigated \cite{17,19,21,22,24,26,30} to compensate for the actuator faults. However, these faults are treated as unintentionally caused natural ones and are required to be bounded. To the best of our knowledge, there is no existing result in the literature to address the resilient control problem of heterogeneous MAS subject to both unknown unbounded sensor and actuator attacks. This is due to the fact that existing methods generally require intact state measurements to construct the local resilient controller against actuator faults. It is more practical yet challenging to devise advanced control protocols incorporating resilience against both unbounded sensor and actuator attacks.

This paper studies the resilient containment control problem of heterogeneous MAS against correlated and unbounded sensor attacks, as well as generally unbounded actuator attacks. These unknown unbounded attack injections could destabilize the dynamical systems and jeopardize the cooperative containment performance. Compared with the alternative results in the existing literature, this paper has the following distinctive characteristics.

$\bullet$ We propose a distributed control framework incorporating resilience against simultaneously unbounded sensor and actuator attacks for heterogeneous MAS. This is more practical in real-world security applications than the existing literature. For example, \cite{17,19,21,22,24,26} address only bounded actuator faults. The intact sensor measurements are needed in the local controller design. The simultaneous sensor and actuator attacks are considered in \cite{20} and \cite{31} for a linear dynamical system and homogeneous MAS, respectively. The simultaneous sensor faults and adversarial attacks are addressed in \cite{32} for heterogeneous MAS. However, the attack signals in \cite{20,31,32} are assumed to be bounded. The resilient consensus problem is studied in \cite{25} for homogeneous MAS against both sensor and actuator faults. Although the sensor faults are relaxed to be unbounded, the system dynamics $A$ is restricted to be nonsingular.

$\bullet$ Rigorous proofs using Lyapunov-based techniques show that the proposed attack-resilient control framework guarantees that the heterogeneous MAS remain stable and the uniformly ultimately bounded (UUB) containment performance is maintained. Although we take the containment control problem as a specific cooperative application in this paper, the proposed method could be applied to many other resilient consensus problems against unbounded sensor and actuator attacks. For example, leaderless and leader-follower consensus, formation control, and so on.

The rest of this paper is organized as follows. The attack-resilient containment control problem of general linear heterogeneous MAS is formulated in Section II. Section III presents the main results to design the distributed attack-resilient controller. In Section IV, an illustrative simulation example validates the efficacy of the proposed results. The conclusions are drawn in Section V.



\section{Preliminaries and Problem Statement}
In this section, we first give the preliminaries in notations and graph theory, then we formulate the attack-resilient containment control problem for heterogeneous MAS.

\subsection{Notations and Graph Theory}
Notations: We use ${\sigma _{\min }}(\cdot)$ and ${\sigma _{\max }}(\cdot)$ to denote the minimum and maximum singular values of a given matrix, respectively. ${I_N} \in {\mathbb{R}^{N \times N}}$ is the identity matrix. ${{\mathbf{1}}_N} \in {\mathbb{R}^N}$ is a vector with all entries of one. $\left\|  \cdot  \right\|$ is the Euclidean norm of a vector. Kronecker product is denoted by $ \otimes $. The operator $\operatorname{diag} \{  \cdot  \}$ builds a block diagonal matrix from its arguments.

Graph theory: We consider the cooperative MAS on a time-invariant communication digraph $\mathscr{G}$, consisting of $N$ followers and $M$ leaders. We use $\mathscr{F}$ and $\mathscr{L}$ to represent the sets of the followers and the leaders, respectively. The interactions among the followers are represented by connected subgraph ${\mathscr{G}_f}$ with the associated adjacency matrix ${\mathcal A} = [{a_{ij}}]$, where ${a_{ij}}$ is the edge weight. $a_{ij}> 0$ if there is an edge from node $j$ to $i$, and node $j$ is called the neighbor of node $i$, otherwise $a_{ij}= 0$. The set of neighbors of node $i$ is denoted as $\mathcal{N}_i$. Define the in-degree matrix as $\mathcal{D} = \operatorname{diag}(d_i) \in {\mathbb{R}^{N \times N}}$ with $d_i = \sum\nolimits_{j \in \mathcal{N}_i} {{a_{ij}}} >0$ and the Laplacian matrix as $\mathcal{L} = \mathcal{D} - \mathcal{A}$. ${\mathcal{G}_k} = \operatorname{diag}\left( {g_{ik}} \right)\in {\mathbb{R}^{N \times N}}$ is the diagonal matrix of pinning gains from the $k^{th}$ leader to each follower.

\subsection{Problem Formulation}
We consider a group of heterogeneous followers with the system dynamics given by
\begin{equation}
{{\dot x}_i}\left( t \right) = {A_i}{x_i}\left( t \right) + {B_i}{{\bar u}_i}\left( t \right),\quad i \in \mathscr{F}
\label{eq1}
\end{equation}
where ${x_i}\left( t \right) \in {\mathbb{R}^n}$ and ${\bar u_i}\left( t \right) \in {\mathbb{R}^{m_i}}$ are the true state and corrupted input of the $i^{th}$ follower, respectively. The local input is under unknown and unbounded actuator attack described by
\begin{equation}
{{\bar u}_i}\left( t \right) = {u_i}\left( t \right) + {\delta _i^a}\left( t \right),
\label{eq2}
\end{equation}
where $u_i\left( t \right)$ is the actual input, and ${\delta _i^a}\left( t \right)$ captures the unknown unbounded actuator attack signal injected to the $i^{th}$ follower. The compromised state measurement of the $i^{th}$ follower is given as
\begin{equation}
{{\bar x}_i}\left( t \right) = {x_i}\left( t \right) + {\delta _i^s}\left( t \right),
\label{eq3}
\end{equation}
where ${\delta _i^s}\left( t \right)$ captures the unknown unbounded sensor attack signal injected to the $i^{th}$ follower. To bypass certain attack detection and identification method on the sensor measurements, instead of launching independently and arbitrarily large bad measurements which are easily detected, the intelligent attackers may inject simultaneous and correlated sensor attacks to the followers \cite{8,11,12}. That is, the sensor attack at $i^{th}$ follower is highly correlated with that of the neighboring followers, i.e.,
\begin{equation}
\delta _i^s\left( t \right) - \delta _j^s\left( t \right) = {\omega_{ij}}\left( t \right),\quad j \in \mathcal{N}_i
\label{eq4}
\end{equation}
where ${\omega_{ij}}\left( t \right)$ is a uniform continuous function which satisfies $\left\| {{\omega _{ij}}(t)} \right\| \le {{\bar \omega }_{ij}}$, $t \ge 0$, with unknown bounds ${{\bar \omega }_{ij}}$. 

\textbf{\textit{Remark} 1:}
While the local sensor attack, $\delta _i^s\left( t \right)$, may be unbounded, the discrepancy of the sensor attack injections between neighboring followers is bounded. By simultaneously launching such correlated and unbounded sensor attacks, the malicious attackers aim to bypass normal attack-detection methods and jeopardize the system stability as well as the desired synchronization performance. For example, in satellite tracking control following an unbounded ramp signal, the attack-detection defensive mechanism may not be triggered in the presence of correlated sensor attack injections.


We consider the following dynamics for the leaders
\begin{equation}
{{\dot x}_k}\left( t \right) = S{x_k}\left( t \right),\quad k \in \mathscr{L}
\label{eq5}
\end{equation}
where ${x _k}\left( t \right) \in {\mathbb{R}^n}$ is the state of the $k^{th}$ leader.

Without loss of any generality, the following standard assumptions are made.

\textbf{\textit{Assumption} 1:}
There exists at least one leader that has a directed path to each follower.

\textbf{\textit{Assumption} 2:} 
The real parts of the eigenvalues of $S$ are non-negative.

\textbf{\textit{Assumption} 3:} 
$({A_i},{B_i})$ is stabilizable for each follower.

\textbf{\textit{Assumption} 4:}
The following linear matrix equation has a solution ${\Gamma _i}$ for each follower
\begin{equation}
S = {A_i} + {B_i}{\Gamma _i}.
\label{eq6}
\end{equation}

\textbf{\textit{Assumption} 5:}
${\dot{\delta}_i^s}\left( t \right)$ and ${\dot{\delta}_i^a}\left( t \right)$ are bounded.

\textbf{\textit{Remark} 2:}
Assumptions 1 and 3 are standard for containment control of heterogeneous MAS with multiple leaders. Assumption 2 is made to avoid the trivial case when the leaders' dynamics are stable and loses no generality. The components of the leaders' states corresponding to the modes associated with the eigenvalues of $S$ with negative real parts will exponentially decay to zero and will in no way affect the asymptotic behavior of the closed-loop system. \eqref{eq6} is the standard regulator equation \cite{2}, when all states $x_i\left( t \right)$ have the same dimension $n$. Assumption 4 holds when $({A_i},{B_i})$ is controllable. We consider the continuously differentiable attack signals ${{\delta}_i^s}\left( t \right)$ and ${{\delta}_i^a}\left( t \right)$. The bounded derivative requirement for the attack signals is standard in the literature. 


We first define the global communication graph topology matrix ${\Psi _k} =( \frac{1}{M}\mathcal{L} + {\mathcal{G}_k})$. Then, we define the following global containment error vector
\begin{equation}
\begin{gathered}
  e\left( t \right)  = x\left( t \right) - {\left( {\sum\limits_{r \in \mathscr{L}} {\left( {{\Psi _r} \otimes {I_n}} \right)} } \right)^{ - 1}} \times  \hfill \\
  \quad \quad    \sum\limits_{k \in \mathscr{L}} {\left( {{\Psi _k} \otimes {I_n}} \right)} \left( {{{\mathbf{1}}_N} \otimes {x_k}\left( t \right)} \right), \hfill \\
\end{gathered}
\label{eq7}
\end{equation}
where $x\left( t \right)= {[ {x_1\left( t \right)^T,...,x_N\left( t \right)^T} ]^T}$.

The containment control objective is to make the trajectory of each local follower converge into the dynamic convex hull spanned by the trajectories of multiple leaders. The following technical results are needed.

\textbf{\textit{Lemma} 1 (\cite{33}):} $\sum\nolimits_{k \in \mathscr{L}} {{\Psi _k}} $ is positive-definite and non-singular if Assumption 1 holds.

\textbf{\textit{Lemma} 2 (\cite{34}):} If Assumption 1 holds, then the containment control objective is achieved if $\mathop {\lim }\limits_{t \to \infty } e \left( t \right) = 0$.

Next, we analyze the susceptibility of the conventional containment control protocols against unbounded sensor and actuator attacks. Consider the following conventional containment control protocols from \cite{33} using state-feedback design
\begin{align}
& {u_i}\left( t \right) = {K_i}{{\bar x}_i}\left( t \right) + {H_i}{\xi _i}\left( t \right), \label{eq8}\\
& \begin{array}{l}
{{\dot \xi }_i}\left( t \right) = S{\xi _i}\left( t \right) + c \times \\
\left( {\sum\limits_{j \in {\mathscr F}} {{a_{ij}}\left( {{\xi _j}\left( t \right) - {\xi _i}\left( t \right)} \right)}  + \sum\limits_{k \in {\mathscr L}} {{g_{ik}}\left( {{x_k}\left( t \right) - {\xi _i}\left( t \right)} \right)} } \right),
\end{array} \label{eq9}
\end{align}
where ${\xi _i}\left( t \right) \in {\mathbb{R}^{n}}$ denotes the state of the dynamic compensator, $c>0$ is some sufficiently large constant, and $K_i, H_i\in {\mathbb{R}^{m_i \times n}}$ are controller gain matrices to be designed later. Combing \eqref{eq1} and \eqref{eq8} yields
\begin{equation}
\begin{array}{l}
{{\dot x}_i}\left( t \right) = \left( {{A_i} + {B_i}{K_i}} \right){x_i}\left( t \right) + {B_i}{H_i}{\xi _i}\left( t \right)\\
\quad \quad  + {B_i}{K_i}\delta _i^s\left( t \right) + {B_i}\delta _i^a\left( t \right),
\end{array}
\label{eq10}
\end{equation}

Since both $\delta _i^s\left( t \right) $ and $\delta _i^a\left( t \right) $ are unbounded, the closed-loop dynamical system \eqref{eq10} becomes unstable using the conventional containment control method. To address such an issue, we introduce the distributed attack-resilient containment control problem, which is precisely stated in Problem 1.

The following convergence result is needed.

\textbf{\textit{Definition} 1 (\cite{35}):} 
The signal $y(t)$ is said to be UUB with ultimate bound $b$ if there exist positive constants $b$ and $e$, independent of ${t_0} \geq 0$, and for every $a \in \left( {0,e} \right)$, there is $T = T\left( {a,b} \right) \geq 0$, independent of $t_0$, such that
\begin{equation}
\left\| {y\left( {{t_0}} \right)} \right\| \leq a\;\; \Rightarrow \;\;\left\| {y\left( t \right)} \right\| \leq b,\;\; \forall t \geq {t_0} + T.
\label{eq11}
\end{equation}

\textbf{\textit{Problem} 1:}
Consider the general linear heterogeneous MAS given by \eqref{eq1} and \eqref{eq5}, in the face of the general unbounded actuator attacks \eqref{eq2} and the correlated and unbounded sensor attacks described in \eqref{eq3} and \eqref{eq4}, the attack-resilient containment control problem is to devise local distributed input $u_i\left( t \right) $ in \eqref{eq2}, such that $e\left( t \right) $ in \eqref{eq7} is UUB, namely, the trajectory of each follower converges to a small neighborhood around the dynamic convex hull spanned by the trajectories of multiple leaders.
\hfill\(\blacksquare\) 

\section{Distributed Attack-Resilient Controller Design}

In this section, we design a novel distributed attack-resilient control architecture for heterogeneous MAS which guarantees the stability of the overall dynamical system and the UUB cooperative containment convergence. To this end, the following measurable error term is first given
\begin{equation}
{\psi_i} \left( t \right) = {\bar x_i} \left( t \right) - {\hat x_i}\left( t \right) - {\hat \delta _i^s}\left( t \right),
\label{eq12}
\end{equation}
where ${{\hat x}_i}\left( t \right)$ and ${{\hat \delta }_i^s}\left( t \right)$ are the estimations of the uncorrupted state $x_i\left( t \right)$ and the sensor attack ${\delta }_i^s\left( t \right)$, respectively. To cope with the general unbounded actuator attacks and the correlated unbounded sensor attacks, combing the dynamic compensator \eqref{eq9} as local observers for the leaders' states, we present the following distributed resilient control structure consisting of the local control protocols, state observers, and sensor attack observers
\begin{align}
& {u_i}\left( t \right) = {K_i}{{\hat x}_i}\left( t \right) + {H_i}{\xi _i}\left( t \right) - \hat \delta _i^a\left( t \right), \label{eq13}\\
& \begin{gathered}
  {{\dot {\hat x}}_i}\left( t \right) = {A_i}{{\hat x}_i}\left( t \right) + {B_i}{K_i}{{\hat x}_i}\left( t \right) + {B_i}{H_i}{\xi _i}\left( t \right) \hfill \\
  \quad \quad \quad  - {B_i}{K_i}{\psi _i}\left( t \right), \hfill \\ 
\end{gathered}  \label{eq14}\\
& \dot {\hat \delta} _i^s \left( t \right) =  -{d_i} {{\hat \delta }_i^s}\left( t \right) + \sum\limits_{j \in \mathscr{F}} {{a_{ij}}\left( {{{\bar x}_j}\left( t \right) - {{\hat x}_j}\left( t \right)} \right)}, \label{eq17}
\end{align}
where ${\hat \delta}_i^a\left( t \right)$ is the estimation of the actuator attack $\delta_i^a\left( t \right)$, and is designed as
\begin{equation}
\hat \delta _i^a\left( t \right) = \frac{{B_i^T{P_i}{\psi _i}\left( t \right){\chi _i}{{\left( t \right)}^2}}}{{\left\| {\psi _i^T\left( t \right){P_i}{B_i}} \right\|{\chi _i}\left( t \right) + {\rho _i}\left( t \right)}},
\label{eq15}
\end{equation}
where ${{\rho _i}\left( t \right)}$ is a uniform continuous function satisfying $\mathop {\lim }\limits_{T \to  + \infty } \int_{{t_0}}^T {{\rho _i}(s)\operatorname{d} s}  \leq {{\bar \rho }_i} <  + \infty$, and ${\chi _i}\left( t \right)$ is an updating parameter using the following adaptive tuning law
\begin{equation}
{{\dot \chi }_i}\left( t \right) = {\mu _i}\left\| {\psi _i^T\left( t \right){P_i}{B_i}} \right\|,
\label{eq16}
\end{equation}
where ${\mu _i}$ is a given positive constant. Given certain symmetric positive-definite matrices $R_i$ and $Q_i$, the control gain matrices $K_i$ and $H_i$ are designed as 
\begin{align}
& {K_i} =  - R_i^{ - 1}B_i^T{P_i},
\label{eq18}\\
& H_i=\Gamma_i-K_i,
\label{eq19}
\end{align}
where $P_i$ is the solution to 
\begin{equation}
A_i^T P_i+P_i A_i+Q_i-P_iB_i{R_i^{-1}}{B_i^T}{P_i}=0.
\label{eq20}
\end{equation}

Similar to \eqref{eq7}, we define the global containment error of the state observer \eqref{eq14} as
\begin{equation}
\begin{gathered}
  \hat e\left( t \right) = \hat x\left( t \right) - {\left( {\sum\nolimits_{r \in \mathscr{L}} {\left( {{\Psi _r} \otimes {I_n}} \right)} } \right)^{ - 1}} \times  \hfill \\
  \quad \quad \sum\limits_{k \in \mathscr{L}} {\left( {{\Psi _k} \otimes {I_n}} \right)} \left( {{{\mathbf{1}}_N} \otimes {x_k}\left( t \right)} \right), \hfill \\ 
\end{gathered}
\label{eq21}
\end{equation}
where $\hat x \left( t \right)= {[ {\hat x_1 \left( t \right)^T,...,\hat x_N \left( t \right)^T} ]^T}$. For convenience, we denote
\begin{equation}
\begin{gathered}
  {{\vec A}_i} = \left[ {\begin{array}{*{20}{c}}
  {{A_i}}&{ - \left( {{A_i} + {d_i}{I_n}} \right)} \\ 
  0&{ - {d_i}{I_n}} 
\end{array}} \right],{{\vec B}_i} = \left[ {\begin{array}{*{20}{c}}
  {{B_i}} \\ 
  0 
\end{array}} \right], \hfill \\
  E = \left[ {\begin{array}{*{20}{c}}
  { - {I_n}} \\ 
  { - {I_n}} 
\end{array}} \right],\vec C = \left[ {\begin{array}{*{20}{c}}
  {{I_n}}&{ - {I_n}} 
\end{array}} \right],{{\vec K}_i} = \left[ {\begin{array}{*{20}{c}}
  {{K_i}}&0 
\end{array}} \right]. \hfill \\ 
\end{gathered}
\label{eq28}
\end{equation}

The following technical result is needed for the next main result.

\textbf{\textit{Lemma} 3:}
Given certain symmetric positive-definite matrix ${M_i} \in {\mathbb R^{n \times n}}$, ${{\vec Q}_i} = \left[ {\begin{array}{*{20}{c}}
  {Q_i}&{ {P_i}{A_i}+{d_i}{P_i}} \\ 
  {A_i^T{P_i} + {d_i}{P_i}}&{ 2{d_i}{M_i}} 
\end{array}} \right]$ is positive-definite. Moreover,
\begin{equation}
{{\vec K}_i} =  - R_i^{ - 1}\vec B_i^T{{\vec P}_i},
\label{eq29}
\end{equation}
where ${{\vec P}_i} = \left[ {\begin{array}{*{20}{c}}
  {{P_i}}&0 \\ 
  0&{{M_i}} 
\end{array}} \right]$ is the solution to
\begin{equation}
\vec A_i^T{{\vec P}_i} + {{\vec P}_i}{{\vec A}_i} + {{\vec Q}_i} - {{\vec P}_i}{{\vec B}_i}R_i^{ - 1}\vec B_i^T{{\vec P}_i} = 0.
\label{eq30}
\end{equation}
\textbf{\textit{Proof:}}
From \eqref{eq18}, \eqref{eq20}, and \eqref{eq28}, it is straightforward to verify that \eqref{eq29} and \eqref{eq30} hold. Next, we prove that given certain symmetric positive-definite matrix $M_i$, ${{\vec Q}_i}$ is positive-definite. Let $\left[ {\begin{array}{*{20}{c}}
  \alpha  \\ 
  \beta  
\end{array}} \right] \in {\mathbb{R}^{2n}}$ be any nonzero column vector with $2n$ real numbers. Then we have
\begin{equation}
\begin{gathered}
 - \left[ {\begin{array}{*{20}{c}}
  {{\alpha ^T}}&{{\beta ^T}} 
\end{array}} \right]{{\vec Q}_i}\left[ {\begin{array}{*{20}{c}}
  \alpha  \\ 
  \beta  
\end{array}} \right] \hfill \\
   =  - {\alpha ^T}{Q_i}\alpha  - 2{\alpha ^T}\left( {{P_i}{A_i} + {d_i}{P_i}} \right)\beta  - 2{\beta ^T}{d_i}{M_i}\beta  \hfill \\
   \leq  - {\sigma _{\min }}\left( {{Q_i}} \right){\left\| \alpha  \right\|^2} + 2{\sigma _{\max }}\left( {{P_i}{A_i} + {d_i}{P_i}} \right)\left\| \alpha  \right\|\left\| \beta  \right\| \hfill \\
   - 2{\sigma _{\min }}\left( {{d_i}{M_i}} \right){\left\| \beta  \right\|^2} \hfill \\
   \leq  - {\sigma _{\min }}\left( {{Q_i}} \right){\left( {\left\| \alpha  \right\| - \frac{{{\sigma _{\max }}\left( {{P_i}{A_i} + {d_i}{P_i}} \right)}}{{{\sigma _{\min }}\left( {{Q_i}} \right)}}\left\| \beta  \right\|} \right)^2} \hfill \\
   - \left( {2{\sigma _{\min }}\left( {{d_i}{M_i}} \right) - \frac{{{{\left( {{\sigma _{\max }}\left( {{P_i}{A_i} + {d_i}{P_i}} \right)} \right)}^2}}}{{{\sigma _{\min }}\left( {{Q_i}} \right)}}} \right){\left\| \beta  \right\|^2}. \hfill \\ 
\end{gathered}
\label{eq31}
\end{equation}

Given $d_i$, $A_i$, $Q_i$ and $P_i$, we can pick $M_i$ such that ${\sigma _{\min }}\left( {{M_i}} \right) \ge \frac{{{{\left( {{\sigma _{\max }}\left( {{P_i}{A_i} + {d_i}{P_i}} \right)} \right)}^2}}}{{2{d_i}{\sigma _{\min }}\left( {{Q_i}} \right)}}$. Then, \eqref{eq31} is strictly negative. That is, $-{{\vec Q}_i}$ is negative-definite. Hence, ${{\vec Q}_i}$ is positive-definite. This completes the proof.
\hfill\(\blacksquare\)

Now, we present the main result to solve Problem 1.

\textbf{\textit{Theorem} 1:}
Consider the heterogeneous MAS consisting of \eqref{eq1} and \eqref{eq5}, in the presence of the general unbounded actuator attacks \eqref{eq2} and the correlated unbounded sensor attacks described in \eqref{eq3} and \eqref{eq4}. Under Assumptions 1,2,3,4, and 5, Problem 1 is solved by designing the distributed attack-resilient control framework consisting of the dynamic compensator \eqref{eq9}, the local control protocols \eqref{eq13}, the state observers \eqref{eq14} and the sensor attack observers \eqref{eq17}.

\textbf{\textit{Proof:}}
To solve Problem 1, we need to prove that $e\left( t \right)$ is UUB using the proposed control framework. Note that $e\left( t \right)=x\left( t \right) - \hat x\left( t \right) + \hat e \left( t \right)= \tilde x\left( t \right) + \hat e\left( t \right)$. In what follows, we first prove the UUB convergence of the followers' states to the observers' states, i.e., $\tilde x\left( t \right)$ is UUB.

Define the estimation error of the local sensor attack as ${\tilde \delta} _i^s\left( t \right) = \delta _i^s \left( t \right)- {{\hat \delta }_i^s}\left( t \right)$. Using \eqref{eq4} and \eqref{eq17} yields 
\begin{equation}
\begin{gathered}
  \dot {\tilde \delta} _i^s \left( t \right)= \dot \delta _i^s\left( t \right) - \dot {\hat \delta} _i^s \left( t \right) 
  \\
  \hfill \\
 =\dot \delta _i^s\left( t \right)  - {d_i}\tilde \delta _i^s \left( t \right)-  \hfill \\
\quad \sum\limits_{j \in \mathscr{F}} {{a_{ij}}\left( {{\psi _j}\left( t \right) - \tilde \delta _j^s\left( t \right)} \right)}  + \sum\limits_{j \in \mathscr{F}} {{a_{ij}}{\omega _{ij}}\left( t \right)}  \hfill \\
   =  - {d_i}\tilde \delta _i^s \left( t \right)- \sum\limits_{j \in \mathscr{F}} {{a_{ij}}\left( {{\psi _j}\left( t \right) - \tilde \delta _j^s\left( t \right)} \right)}  + {\Delta _i}\left( t \right), \hfill \\ 
\end{gathered}
\label{eq32}
\end{equation}
where ${\Delta _i} \left( t \right)= \sum\nolimits_{j \in \mathscr{F}} {{a_{ij}}{\omega _{ij}}\left( t \right)}  + \dot \delta _i^s\left( t \right)$ is bounded under Assumption 5.

Using \eqref{eq12}, \eqref{eq13}, \eqref{eq14}, and \eqref{eq32} yields
\begin{equation}
\begin{gathered}
  {{\dot \psi }_i}\left( t \right) = {{\dot {\bar x}}_i}\left( t \right) - {{\dot {\hat x}}_i}\left( t \right) - \dot {\hat \delta} _i^s \left( t \right) \hfill \\
   = {A_i}{x_i}\left( t \right) + {B_i}\tilde \delta _i^a\left( t \right) - {A_i}{{\hat x}_i}\left( t \right) + {B_i}{K_i}{\psi _i}\left( t \right) \hfill \\
   - {d_i}\tilde \delta _i^s\left( t \right) - \sum\limits_{j \in \mathscr{F}} {{a_{ij}}\left( {{\psi _j}\left( t \right) - \tilde \delta _j^s\left( t \right)} \right)}  + {\Delta _i}\left( t \right) \hfill \\
   = \left( {{A_i} + {B_i}{K_i}} \right){\psi _i}\left( t \right) - \left( {{A_i} + {d_i}{I_n}} \right)\tilde \delta _i^s\left( t \right) + {B_i}\tilde \delta _i^a\left( t \right) \hfill \\
   - \sum\limits_{j \in \mathscr{F}} {{a_{ij}}\left( {{\psi _j}\left( t \right) - \tilde \delta _j^s\left( t \right)} \right)}  + {\Delta _i}\left( t \right). \hfill \\ 
\end{gathered}
\label{eq33}
\end{equation}

Denote ${\vartheta _i}\left( t \right) = \left[ {\begin{array}{*{20}{c}}
  {{\psi _i}}\left( t \right) \\ 
  {\tilde \delta _i^s}\left( t \right) 
\end{array}} \right]$ and use \eqref{eq28}, we then put \eqref{eq32} and \eqref{eq33} in the following compact form 
\begin{equation}
\begin{gathered}
  {{\dot \vartheta }_i}\left( t \right) = \left( {{{\vec A}_i} + {{\vec B}_i}{{\vec K}_i}} \right){\vartheta _i}\left( t \right) + {{\vec B}_i}\tilde \delta _i^a\left( t \right)\hfill \\
  \quad \quad \quad + E\vec C\sum\limits_{j \in \mathscr{F}} {{a_{ij}}{\vartheta _j}\left( t \right)}  - E{\Delta _i}\left( t \right) \hfill \\
  \quad \quad  = \left( {{{\vec A}_i} + {{\vec B}_i}{{\vec K}_i}} \right){\vartheta _i}\left( t \right) + {{\vec B}_i}\tilde \delta _i^a\left( t \right) + {{\vec \Delta }_i}\left( t \right), \hfill \\ 
\end{gathered}
\label{eq34}
\end{equation}
where ${{\vec \Delta }_i}\left( t \right) = E\vec C\sum\nolimits_{j \in \mathscr{F}} {{a_{ij}}{\vartheta _j}\left( t \right)}  - E{\Delta _i}\left( t \right)$, which could be considered as a bounded term.

Under Assumption 3, it follows from Lemma 3 that ${{\vec P}_i}$ is symmetric positive-definite. Hence, there exists a nonsingular matrix ${\Phi _i}$ satisfying ${{\vec P}_i} = \Phi _i^T{\Phi _i}$. Consider the following Lyapunov function candidate
\begin{equation}
{V_i} \left( t \right)= {\left( {{\Phi _i}{\vartheta _i}\left( t \right)} \right)^T}{\Phi _i}{\vartheta _i}\left( t \right),
\label{eq35}
\end{equation}
with its time derivative given as
\begin{equation}
{{\dot V}_i} \left( t \right)= 2{\vartheta _i}\left( t \right)^T\Phi _i^T{\Phi _i}{{\dot \vartheta }_i}\left( t \right) = 2\vartheta _i\left( t \right)^T{{\vec P}_i}{{\dot \vartheta }_i}\left( t \right).
\label{eq36}
\end{equation}


Let ${{A'_i}} = {{\vec A}_i} + {{\vec B}_i}{{\vec K}_i}$ and ${Q'_i}={\vec Q_i} + \vec K_i^T{R_i}{\vec K_i}$. Note that ${Q'_i}$ is positive-definite. Substituting \eqref{eq29} into \eqref{eq30} yields
\begin{equation}
{A'_i}^T{\vec P_i} + {\vec P_i}{A'_i} + {Q'_i} = 0.
\label{eq37}
\end{equation}

Substituting \eqref{eq34} into \eqref{eq36} yields
\begin{equation}
\begin{array}{l}
{{\dot V}_i}\left( t \right) = 2{\vartheta _i}\left( t \right)^T{{\vec P}_i}\left( {{{A'}_i}{\vartheta _i}\left( t \right) + {{\vec B}_i}\tilde \delta _i^a\left( t \right) + {{\vec \Delta }_i}\left( t \right)} \right)\\
 =  - \vartheta _i\left( t \right)^T{{Q'}_i}{\vartheta _i}\left( t \right) + 2{\vartheta _i}\left( t \right)^T{{\vec P}_i}\left( {{{\vec B}_i}\tilde \delta _i^a\left( t \right) + {{\vec \Delta }_i}\left( t \right)} \right)\\
\le  - {\sigma _{\min }}\left( {{{Q'}_i}} \right){\left\| {{\vartheta _i}}\left( t \right) \right\|^2} + 2{\sigma _{\max }}\left( {{{\vec P}_i}} \right)\left\| {{{\vec \Delta }_i}}\left( t \right) \right\|\left\| {{\vartheta _i}}\left( t \right) \right\|\\
\quad \quad  + 2{\vartheta _i}\left( t \right)^T{{\vec P}_i}{{\vec B}_i}\tilde \delta _i^a\left( t \right).
\end{array}
\label{eq38}
\end{equation}

Note that $B_i^T{P_i}{\psi _i}\left( t \right) = \vec B_i^T{{\vec P}_i}{\vartheta _i}\left( t \right)$. Hence \eqref{eq15} and \eqref{eq16} can be rewritten as 
\begin{equation}
\hat \delta _i^a \left( t \right)= \frac{{\vec B_i^T{{\vec P}_i}{\vartheta _i}\left( t \right)\chi _i\left( t \right)^2}}{{\left\| {\vartheta _i\left( t \right)^T{{\vec P}_i}{{\vec B}_i}} \right\|{\chi _i}\left( t \right) + {\rho _i}\left( t \right)}},
\label{eq39}
\end{equation}
\begin{equation}
{{\dot \chi }_i}\left( t \right) = {\mu_i}\left\| {\vartheta _i\left( t \right)^T{{\vec P}_i}{{\vec B}_i}} \right\|.
\label{eq40}
\end{equation}

Using \eqref{eq39}, we obtain 
\begin{equation}
\begin{array}{l}
{\vartheta _i}\left( t \right)^T{{\vec P}_i}{{\vec B}_i}\tilde \delta _i^a \left( t \right)\hfill \\
   = {\vartheta _i}\left( t \right)^T{{\vec P}_i}{{\vec B}_i}\delta _i^a \left( t \right)- \frac{{{{\left\| {\vartheta _i\left( t \right)^T{{\vec P}_i}{{\vec B}_i}} \right\|}^2}\chi _i\left( t \right)^2}}{{\left\| {\vartheta _i\left( t \right)^T{{\vec P}_i}{{\vec B}_i}} \right\|{\chi _i}\left( t \right) + {\rho _i}\left( t \right)}}\hfill \\
 \le \left\| {{\vartheta _i}\left( t \right)^T{{\vec P}_i}{{\vec B}_i}} \right\|\left\| {\delta _i^a}\left( t \right) \right\| - \frac{{{{\left\| {\vartheta _i\left( t \right)^T{{\vec P}_i}{{\vec B}_i}} \right\|}^2}\chi _i\left( t \right)^2}}{{\left\| {\vartheta _i\left( t \right)^T{{\vec P}_i}{{\vec B}_i}} \right\|{\chi _i}\left( t \right) + {\rho _i}\left( t \right)}}\hfill \\
 \le \frac{{\left\| {{\vartheta _i}\left( t \right)^T{{\vec P}_i}{{\vec B}_i}} \right\|{\chi _i}\left( t \right)\left( {\left\| {\delta _i^a}\left( t \right) \right\| - {\chi _i}\left( t \right)} \right) + \left\| {\delta _i^a}\left( t \right) \right\|{\rho _i}\left( t \right)}}{{{\chi _i}\left( t \right) + \left( {{{{\rho _i}\left( t \right)} \mathord{\left/
 {\vphantom {{{\rho _i}\left( t \right)} {\left\| {{\vartheta _i}\left( t \right)^T{{\vec P}_i}{{\vec B}_i}} \right\|}}} \right.
 \kern-\nulldelimiterspace} {\left\| {{\vartheta _i}\left( t \right)^T{{\vec P}_i}{{\vec B}_i}} \right\|}}} \right)}}.
\end{array}
\label{eq41}
\end{equation}

Since $\frac{{{\mathop{\operatorname{d}}\nolimits} \left\| {\delta _i^a} \right\|}}{{{\mathop{\operatorname {d}}\nolimits} t}} = \frac{{\delta {{_i^a}^T}\dot \delta _i^a}}{{\left\| {\delta _i^a} \right\|}} \le \left\| {\dot \delta _i^a} \right\|$ and ${\dot \delta _i^a}$ is bounded under Assumption 5, $\frac{{{\mathop{\operatorname{d}}\nolimits} \left\| {\delta _i^a} \right\|}}{{{\mathop{\operatorname {d}}\nolimits} t}}$ is also bounded. Choosing $\left\| {{\vartheta _i}{{\left( t \right)}^T}{{\vec P}_i}{{\vec B}_i}} \right\| > \frac{1}{{{\mu _i}}}\frac{{{\mathop{\operatorname d}\nolimits} \left\| {\delta _i^a} \right\|}}{{{\mathop{\operatorname d}\nolimits} t}}$ and using \eqref{eq40}, we obtain that ${{\dot \chi }_i} > \frac{{{\mathop{\operatorname d}\nolimits} \left\| {\delta _i^a} \right\|}}{{{\mathop{\operatorname d}\nolimits} t}}$. Since $\frac{{{\mathop{\operatorname{d}}\nolimits} \left\| {\delta _i^a} \right\|}}{{{\mathop{\operatorname {d}}\nolimits} t}}$ is bounded and $\mathop {\lim }\limits_{T \to  + \infty } \int_{{t_0}}^T {{\rho _i}(s)\operatorname{d} s}  \leq {{\bar \rho }_i} <  + \infty$, we have $\mathop {\lim }\limits_{t \to  + \infty } \left\| {\delta _i^a\left( t \right)} \right\|{\rho _i}\left( t \right) = 0$. Hence, there exists $\tau  > 0$, such that
\begin{equation}
{\vartheta _i}\left( t \right)^T{{\vec P}_i}{{\vec B}_i}\tilde \delta _i^a \left( t \right) \leq 0,\quad \forall t \geq \tau .
\label{eq42}
\end{equation}

Substituting \eqref{eq42} into \eqref{eq38} yields 
\begin{equation}
\begin{array}{l}
{{\dot V}_i}\left( t \right) \le  - {\sigma _{\min }}\left( {{Q'_i}} \right){\left\| {{\vartheta _i}\left( t \right)} \right\|^2} \hfill \\
\quad \quad \quad + 2{\sigma _{\max }}\left( {{{\vec P}_i}} \right)\left\| {{{\vec \Delta }_i}\left( t \right)} \right\|\left\| {{\vartheta _i}\left( t \right)} \right\| \hfill \\
\quad \quad  \le  - {\sigma _{\min }}\left( {{Q'_i}} \right)\left\| {{\vartheta _i}\left( t \right)} \right\| \times \hfill \\
\quad \quad \quad \left( {\left\| {{\vartheta _i}\left( t \right)} \right\| - \frac{{2{\sigma _{\max }}\left( {{{\vec P}_i}} \right)\left\| {{{\vec \Delta }_i}\left( t \right)} \right\|}}{{{\sigma _{\min }}\left( {{Q'_i}} \right)}}} \right), \; \forall t \ge \tau 
\end{array}
\label{eq43}
\end{equation}

Then, by choosing ${\left\| {{\vartheta _i}\left( t \right)} \right\| \ge \frac{{2{\sigma _{\max }}\left( {{{\vec P}_i}} \right)\left\| {{{\vec \Delta }_i}\left( t \right)} \right\|}}{{{\sigma _{\min }}\left( {{{Q'}_i}} \right)}}}$, we obtain
\begin{equation}
{{\dot V}_i}\left( t \right) \leq 0,\quad \forall t \geq \tau
\label{eq44}
\end{equation}

That is, ${{\vartheta _i}}\left( t \right)$ is UUB \cite{36}. Since ${\vartheta _i}\left( t \right) = \left[ {\begin{array}{*{20}{c}}
  {{\psi _i}\left( t \right)} \\ 
  {\tilde \delta _i^s}\left( t \right) 
\end{array}} \right]$, ${\psi _i}\left( t \right)$ is bounded. Moreover, ${{\tilde x}_i}\left( t \right)$ is bounded since ${{\tilde x}_i}\left( t \right) = \vec C{\vartheta _i}\left( t \right)$. 

Next we prove that the proposed observers given by \eqref{eq9} and \eqref{eq14} guarantee the UUB convergence of the state estimation, $\hat x\left( t \right)$, to the convex hull spanned by the leaders' states, i.e., $\hat e\left( t \right)$ is UUB.

Define the following global containment error for the compensator state
\begin{equation}
\begin{gathered}
  \eta \left( t \right) = \xi\left( t \right)  - {\left( {\sum\limits_{r \in \mathscr{L}} {\left( {{\Psi _r} \otimes {I_n}} \right)} } \right)^{ - 1}} \times \hfill \\
 \quad \quad \quad  \sum\limits_{k \in \mathscr{L}} {\left( {{\Psi _k} \otimes {I_n}} \right)} \left( {{{\mathbf{1}}_N} \otimes {x_k}\left( t \right)} \right), \hfill \\
\end{gathered} 
\label{eq22}
\end{equation}
and the global error vector $\varepsilon\left( t \right) = \hat x\left( t \right) - \xi\left( t \right) $, where $\xi\left( t \right)={[ {{\xi_1\left( t \right)^T},...,{\xi_N \left( t \right)^T}} ]^T}$. Using \eqref{eq5} and \eqref{eq9} yields
\begin{equation}
\begin{gathered}
  \dot \eta \left( t \right) = \dot \xi \left( t \right) - {\left( {\sum\limits_{r \in \mathscr{L}} {\left( {{\Psi _r} \otimes {I_n}} \right)} } \right)^{ - 1}}\times \hfill \\
 \quad \quad \quad \sum\limits_{k \in \mathscr{L}} {\left( {{\Psi _k} \otimes {I_n}} \right)} \left( {{{\mathbf{1}}_N} \otimes {{\dot x}_k}\left( t \right)} \right) \hfill \\
 = \left( {{I_N} \otimes S} \right)\xi \left( t \right) - c\sum\limits_{k \in \mathscr{L}} {\left( {{\Psi _k} \otimes {I_n}} \right)} \left( {\xi\left( t \right) - {{\mathbf{1}}_N} \otimes {x_k}\left( t \right)} \right)   \hfill \\
 - \left( {{I_N} \otimes S} \right){\left( {\sum\limits_{r \in \mathscr{L}} {\left( {{\Psi _r} \otimes {I_n}} \right)} } \right)^{ - 1}}\sum\limits_{k \in \mathscr{L}} {\left( {{\Psi _k} \otimes {I_n}} \right)} \left( {{{\mathbf{1}}_N} \otimes {x_k}\left( t \right)} \right) \hfill \\
   = \left( {\left( {{I_N} \otimes S} \right) - c\sum\limits_{k \in \mathscr{L}} {\left( {{\Psi _k} \otimes {I_n}} \right)} } \right)\eta \left( t \right) . \hfill \\ 
\end{gathered}
\label{eq23}
\end{equation}

The eigenvalues of ${\left( {{I_N} \otimes S} \right) - c\sum\nolimits_{k \in \mathscr{L}} {\left( {{\Psi _k} \otimes {I_n}} \right)} }$ are $\left\{ {{\lambda _i}\left( S \right) - c{\lambda _j}\left( {\sum\nolimits_{k \in \mathscr{L}} {{\Psi _k}} } \right)} \right\},\forall i = 1,...,n,j = 1,...,N$. Under Assumption 1, it follows from Lemma 1 that the real parts of ${\lambda _j}\left( {\sum\nolimits_{k \in \mathscr{L}} {{\Psi _k}} }\right)$ are positive. Hence, ${\left( {{I_N} \otimes S} \right) - c\sum\nolimits_{k \in \mathscr{L}} {\left( {{\Psi _k} \otimes {I_n}} \right)} }$ is Hurwitz if $c$ is sufficiently large. Then, we obtain $\mathop {\lim }\limits_{t \to \infty } \eta \left( t \right) = 0$. 

Given Assumptions 3 and 4. Using \eqref{eq6}, \eqref{eq9}, \eqref{eq14} and \eqref{eq19} yields
\begin{equation}
\begin{gathered}
  \dot \varepsilon \left( t \right) = \dot {\hat x}\left( t \right) - \dot \xi \left( t \right) \hfill \\
 \; \;  \; = \operatorname{diag} \left( {{A_i} + {B_i}{K_i}} \right)\varepsilon \left( t \right) - \operatorname{diag} \left( {{B_i}{K_i}} \right)\psi \left( t \right) \hfill \\
 \quad \quad  + c\sum\limits_{k \in \mathscr{L}} {\left( {{\Psi _k} \otimes {I_n}} \right)\left( {\xi \left( t \right) - {{\mathbf{1}}_N} \otimes {x_k}\left( t \right)} \right)}  \hfill \\
 \; \; \; = \operatorname{diag} \left( {{A_i} + {B_i}{K_i}} \right)\varepsilon \left( t \right) + c\sum\limits_{k \in \mathscr{L}} {\left( {{\Psi _k} \otimes {I_n}} \right)} \eta \left( t \right) \hfill \\
 \quad \quad  - \operatorname{diag} \left( {{B_i}{K_i}} \right)\psi\left( t \right), \hfill \\
\end{gathered}
\label{eq24}
\end{equation}
where $\psi\left( t \right)={[ {{\psi_1\left( t \right)^T},...,{\psi_N\left( t \right)^T}} ]^T}$. Let $\bar A_i=A_i + B_i K_i$ and $\bar Q_i = Q_i + K_i^T R_i K_i$. Note that $\bar Q_i$ is positive-definite. Substituting \eqref{eq18} into \eqref{eq20} yields ${{\bar A}_i}^T{P_i} + {P_i}{{\bar A}_i}=-{\bar Q_i}$. Hence ${{\bar A}_i}$ is Hurwitz. Note that $\mathop {\lim }\limits_{t \to \infty } \eta \left( t \right) = 0$ and $\psi_i\left( t \right)$ is bounded, we then obtain that $\varepsilon \left( t \right)$ is bounded. Since $\hat e \left( t \right)= \varepsilon \left( t \right) + \eta \left( t \right)$, $\hat e \left( t \right)$ is also bounded. Finally, we obtain that $e\left( t \right)= \tilde x\left( t \right) + \hat e\left( t \right)$ is UUB, i.e., Problem 1 is solved.
\hfill\(\blacksquare\)

\textbf{\textit{Remark} 3:}
In the presence of the correlated and unbounded sensor attacks, instead of using the true state measurement $x_i\left( t \right)$ in the local controller design as shown in most of the existing literature, we use the measurable error term ${\psi_i} \left( t \right) = {\bar x_i} \left( t \right) - {\hat x_i}\left( t \right) - {\hat \delta _i^s}\left( t \right)$ in the local controller design shown in \eqref{eq14}. As seen from Lemma 3 and Theorem 1, this complicates the stability proof of the overall dynamical system as it incorporates an augmented closed-loop error dynamics shown in \eqref{eq34}.

\textbf{\textit{Remark} 4:}
As seen from \eqref{eq23}, the information of the global graph topology described in $\sum\nolimits_{k \in {\mathscr L}} {{\Psi _k}} $ is needed in the choice of the coupling gain $c$. Note that, as long as Assumption 1 holds, we can always choose $c$ sufficiently large, without requiring detailed graph information. Moreover, an alternative method is developed in \cite{37} to adaptively design $c$ in a fully distributed manner and requires no global information.

\textbf{\textit{Remark} 5:}
Compared with some recent research dealing with both bounded sensor and actuator faults \cite{20,25,31,32}, we propose a distributed resilient control framework to incorporate resilience against simultaneously unbounded sensor and actuator attacks for general linear heterogeneous MAS. The proposed results could be applied to many other cooperative control applications.

\section{Simulation Results}
\begin{figure}
\begin{center}
\includegraphics[scale=0.45]{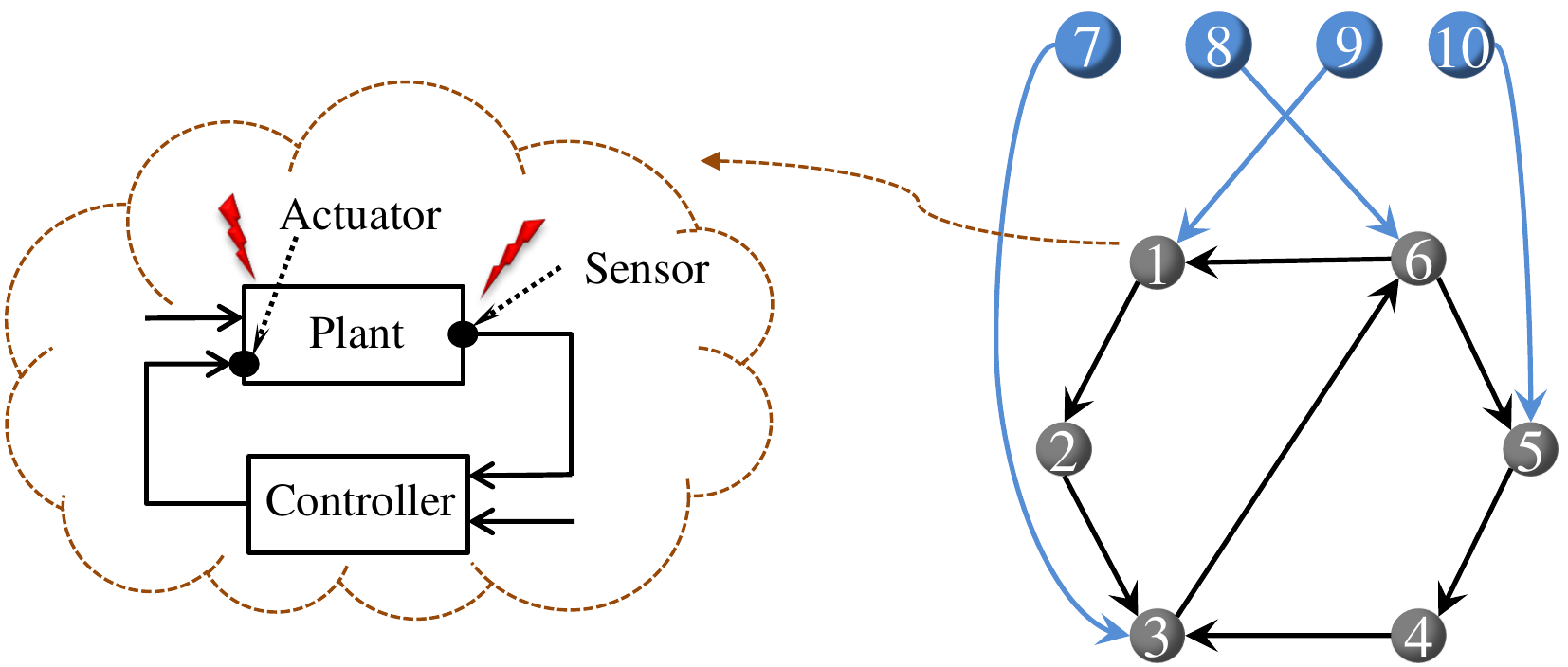}
\caption{The communication digraph $\mathscr{G}$.}
\label{Fig1}
\end{center} 
\end{figure}

In this section, a numerical example is presented to demonstrate the resilience of the proposed distributed control protocols against simultaneously unbounded sensor and actuator attacks. We consider a networked heterogeneous MAS illustrated in Fig.~\ref{Fig1}, consisting of six followers denoted by agents 1, 2,..., 6, and four leaders denoted by agents 7,8,...,10, with their heterogeneous dynamics described by 
\begin{align}
& {{\dot x}_i} = \left[ {\begin{array}{*{20}{c}}
2&{ - 3}\\
4&{ - 4}
\end{array}} \right]{x_i} + \left[ {\begin{array}{*{20}{c}}
1\\
3
\end{array}} \right]{{\bar u}_i},\;\;{\kern 1pt} i = 1,2, \nonumber \\
& {{\dot x}_i} = \left[ {\begin{array}{*{20}{c}}
2&0\\
3&3
\end{array}} \right]{x_i} + \left[ {\begin{array}{*{20}{c}}
{ - 1}\\
{ - 2}
\end{array}} \right]{{\bar u}_i},\;\;{\kern 1pt} i = 3,4,
\nonumber \\
& {{\dot x}_i} = \left[ {\begin{array}{*{20}{c}}
{ - 5}&2\\
4&{ - 3}
\end{array}} \right]{x_i} + \left[ {\begin{array}{*{20}{c}}
2\\
{ - 1}
\end{array}} \right]{{\bar u}_i},\;\;{\kern 1pt} i = 5,6,
\nonumber \\
& {{\dot x}_i} = \left[ {\begin{array}{*{20}{c}}
1&{ - 2}\\
1&{ - 1}
\end{array}} \right]{x_i},\;\;{\kern 1pt} i = 7,8,9,10.
\nonumber
\end{align}

Based on \eqref{eq3} and \eqref{eq4}, we consider the following unbounded and correlated sensor attacks injected to local followers.
\begin{align}
& \delta _1^s\left( t \right) = \left[ {\begin{array}{*{20}{c}}
1\\
{0.5}
\end{array}} \right]t - 0.1\left[ {\begin{array}{*{20}{c}}
{\sin (t)}\\
{\sin (t)}
\end{array}} \right], \nonumber \\
& \delta _2^s\left( t \right) = \left[ {\begin{array}{*{20}{c}}
1\\
{0.5}
\end{array}} \right]t + 0.2\left[ {\begin{array}{*{20}{c}}
{\cos(t)}\\
{ - \cos (t)}
\end{array}} \right],
\nonumber \\
& \delta _3^s\left( t \right) = \left[ {\begin{array}{*{20}{c}}
1\\
{0.5}
\end{array}} \right]t + 0.3\left[ {\begin{array}{*{20}{c}}
{ - \sin (t)}\\
{\cos(t)}
\end{array}} \right],
\nonumber
\end{align}
\begin{align}
& \delta _4^s\left( t \right) = \left[ {\begin{array}{*{20}{c}}
1\\
{0.5}
\end{array}} \right]t + 0.4\left[ {\begin{array}{*{20}{c}}
{\sin(t)}\\
{ - \cos (t)}
\end{array}} \right],
\nonumber \\
& \delta _5^s\left( t \right) = \left[ {\begin{array}{*{20}{c}}
1\\
{0.5}
\end{array}} \right]t - 0.5\left[ {\begin{array}{*{20}{c}}
{\cos(t)}\\
{\sin(t)}
\end{array}} \right],
\nonumber \\
& \delta _6^s\left( t \right) = \left[ {\begin{array}{*{20}{c}}
1\\
{0.5}
\end{array}} \right]t + 0.6\left[ {\begin{array}{*{20}{c}}
{\cos(t)}\\
{ - \sin (t)}
\end{array}} \right].
\nonumber
\end{align}

Based on \eqref{eq2}, we consider the following general unknown unbounded actuator attacks injected to local followers.
\begin{align}
& \delta _1^a\left( t \right) = 0.1t,\quad  \delta _2^a\left( t \right) =  - 0.2t, \quad \delta _3^a\left( t \right) = 0.3t,
\nonumber \\
& \delta _4^a\left( t \right) =  - 0.4t, \quad \delta _5^a\left( t \right) = 0.5t, \quad \delta _6^a\left( t \right) =  - 0.6t.
\nonumber
\end{align}

To show the comparative performance, we run the simulation using the conventional cooperative control method, i.e., \eqref{eq8} and \eqref{eq9}, and the proposed attack-resilient control protocols composed of \eqref{eq9}, \eqref{eq12}, and \eqref{eq13} to \eqref{eq20}. We set $c=10$, $\rho_i\left( t \right)={e^{ - 0.01t}}$, ${\mu _i} = 2$, $R_i=1$, ${Q_i} = \left[ {\begin{array}{*{20}{c}}
3&0\\
0&3
\end{array}} \right]$. By solving \eqref{eq6}, \eqref{eq18}, \eqref{eq19}, and \eqref{eq20}, we obtain
\begin{align}
& {\Gamma _{1,2}} = \left[ {\begin{array}{*{20}{c}}
{ - 1}&1
\end{array}} \right],{\Gamma _{3,4}} = \left[ {\begin{array}{*{20}{c}}
1&2
\end{array}} \right],{\Gamma _{5,6}} = \left[ {\begin{array}{*{20}{c}}
3&{ - 2}
\end{array}} \right], \nonumber \\
& {K _{1,2}} = \left[ {\begin{array}{*{20}{c}}
 {2.158}&{-2.321}
\end{array}} \right],{K_{3,4}} = \left[ {\begin{array}{*{20}{c}}
 {-3.180}&{7.416}
\end{array}} \right], \nonumber \\
& {K _{5,6}} = \left[ {\begin{array}{*{20}{c}}
 {-0.816}&{-0.290}
\end{array}} \right],{H_{1,2}} = \left[ {\begin{array}{*{20}{c}}
 {-3.158}&{3.321}
\end{array}} \right], \nonumber \\
& {H _{3,4}} = \left[ {\begin{array}{*{20}{c}}
 {4.180}&{-5.416}
\end{array}} \right],{H_{5,6}} = \left[ {\begin{array}{*{20}{c}}
 {3.816}&{-1.710}
\end{array}} \right].
\nonumber
\end{align}

\begin{figure}
\begin{center}
\includegraphics[scale=0.6]{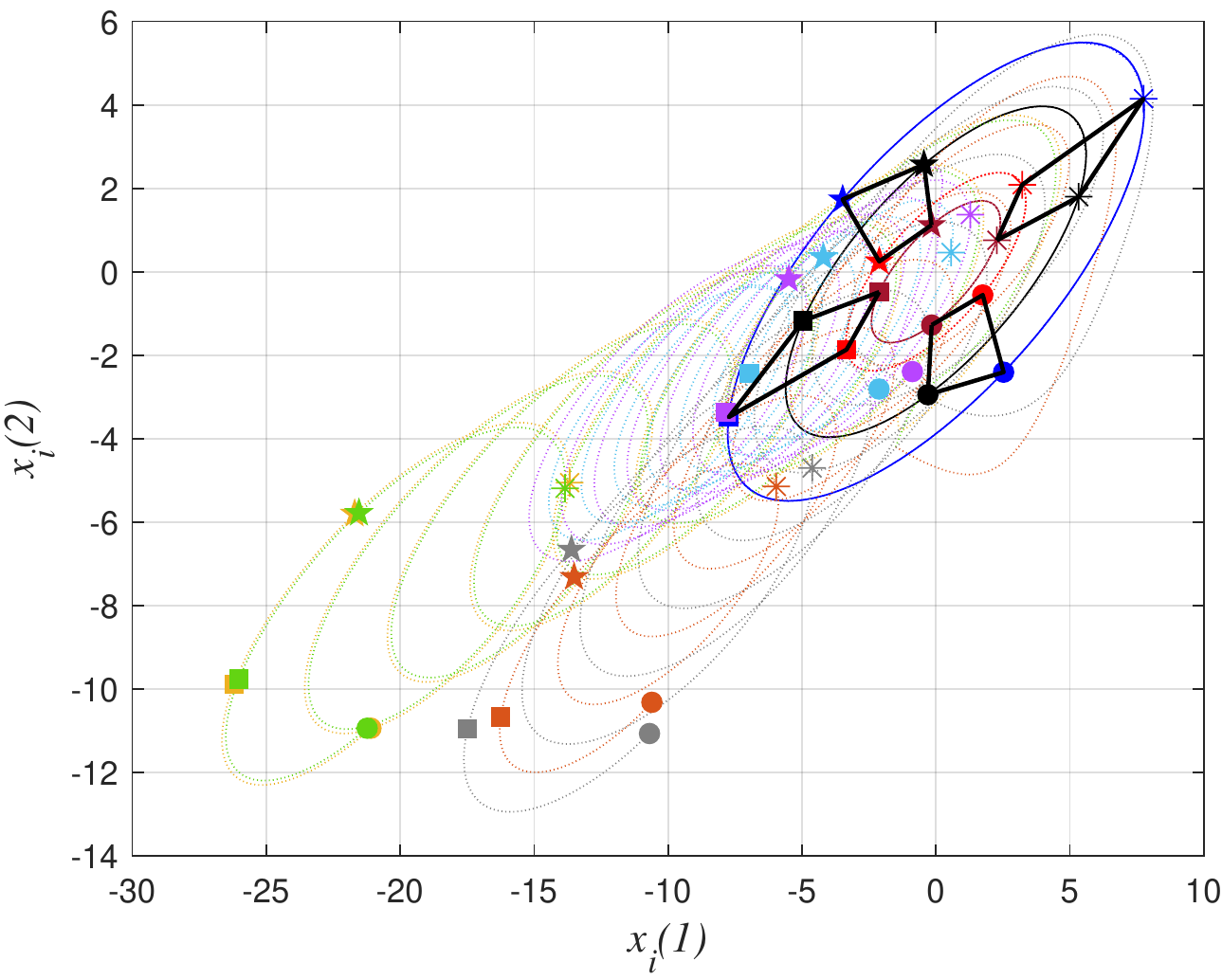}
\caption{Trajectories of the agents under unbounded sensor attacks, using the conventional cooperative control method.}
\label{Fig2}
\end{center} 
\end{figure}

\begin{figure}
\begin{center}
\includegraphics[scale=0.6]{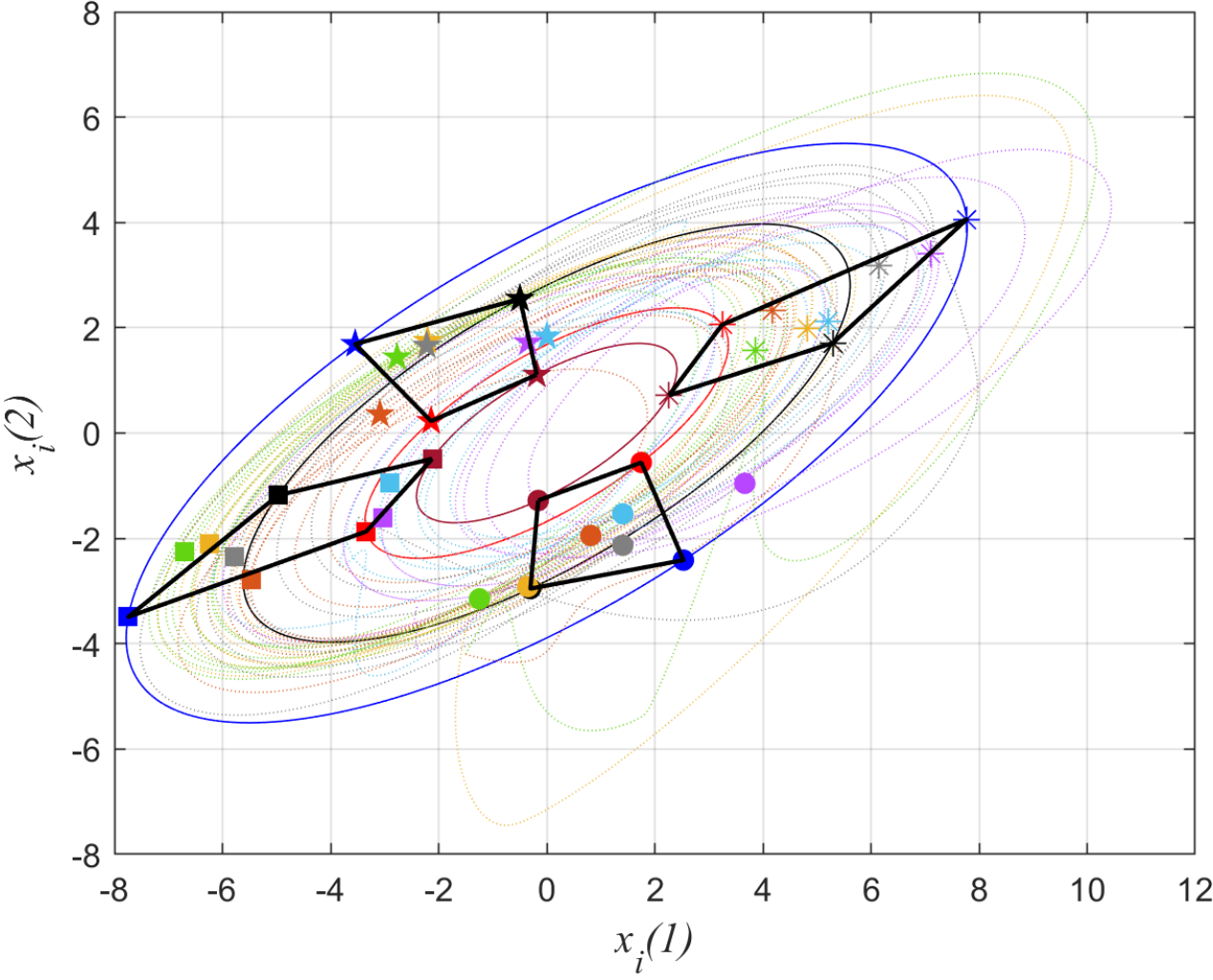}
\caption{Trajectories of the agents under unbounded sensor attacks, using the proposed resilient control method.}
\label{Fig3}
\end{center} 
\end{figure}

We first evaluate the system resilience against correlated and unbounded sensor attacks. Fig.~\ref{Fig2} and Fig.~\ref{Fig3} show the trajectories of all the agents using the conventional and the proposed control methods, respectively, for $t \in \left[ {0,44} \right]s$, where ${x_i} = {[ {\begin{array}{*{20}{c}}
  {{x_i}\left( 1 \right)}&{{x_i}\left( 2 \right)} 
\end{array}}]^T}$. The four leaders move in ellipses. The snapshots of all the agents are taken for four instants $\left( t=39s,41s,42s,44s\right)$ marked with $\ast$, $\bigstar$, $\blacksquare$, and $\bullet$, respectively. The convex hulls formed by the four leaders are illustrated using black solid lines. As seen, under correlated and unbounded sensor attacks, trajectories of the followers diverge and are unbounded, using the conventional method. By contrast, trajectories of the followers, using the proposed resilient approach, stay in a small neighborhood around the convex hull spanned by the leaders, i.e., the UUB convergence result is obtained.

\begin{figure}

\begin{center}
\includegraphics[scale=0.6]{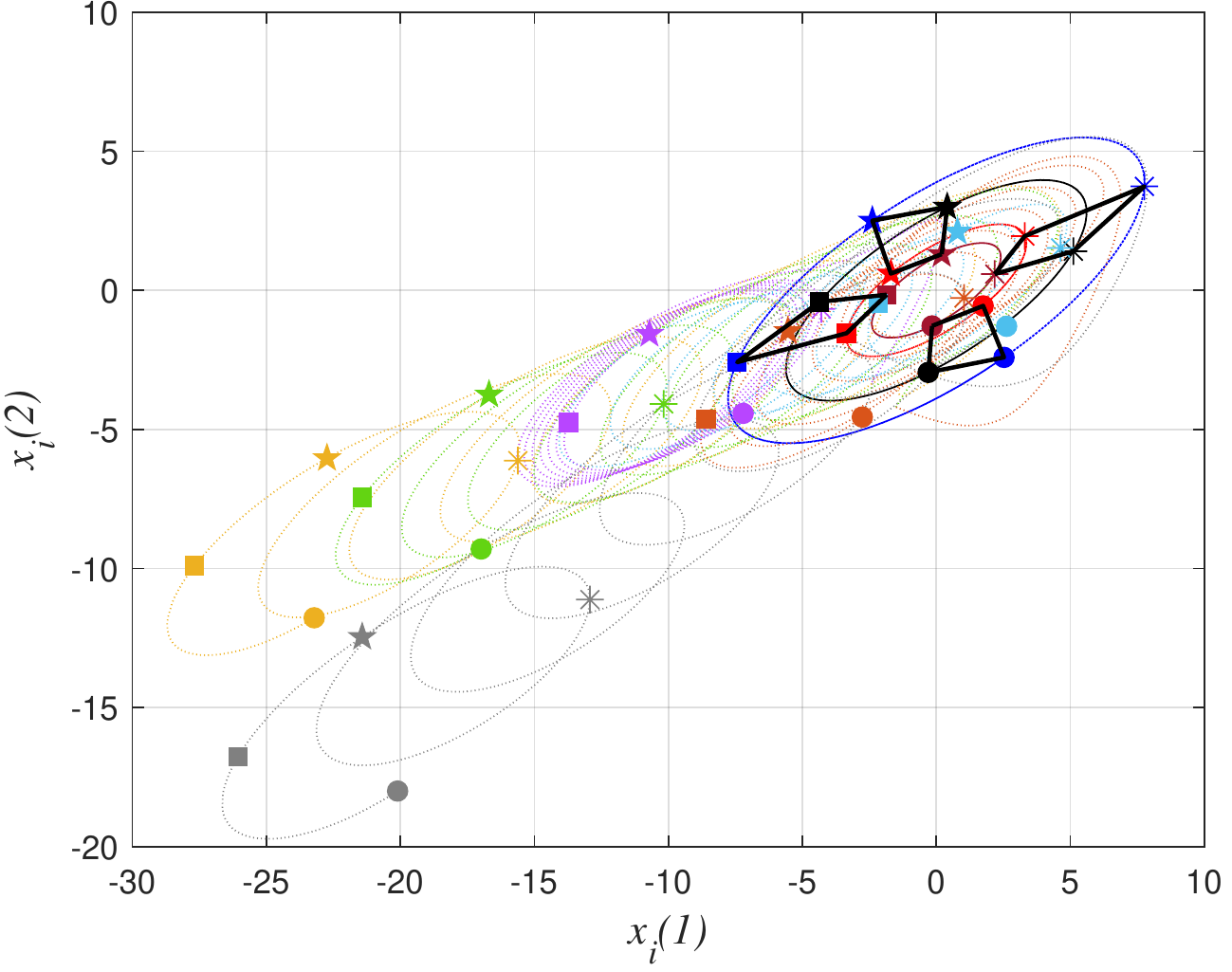}
\caption{Trajectories of the agents under the unbounded sensor and actuator attacks, using the conventional cooperative control method.}
\label{Fig4}
\end{center} 
\end{figure}
\begin{figure}
\begin{center}
\includegraphics[scale=0.6]{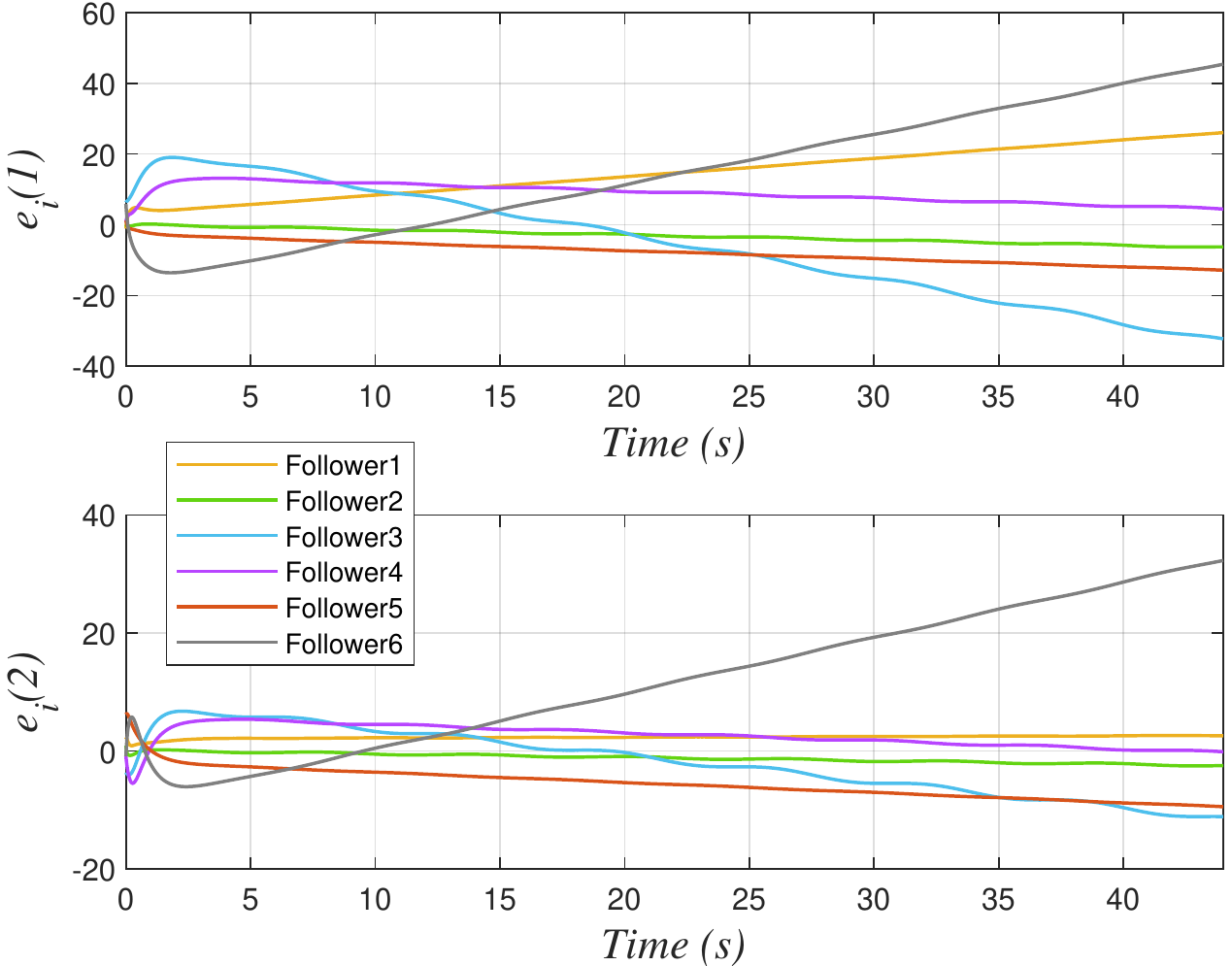}
\caption{Containment errors of the followers over time, using the conventional cooperative control method.}
\label{Fig5}
\end{center} 
\end{figure}

\begin{figure}
\begin{center}
\includegraphics[scale=0.6]{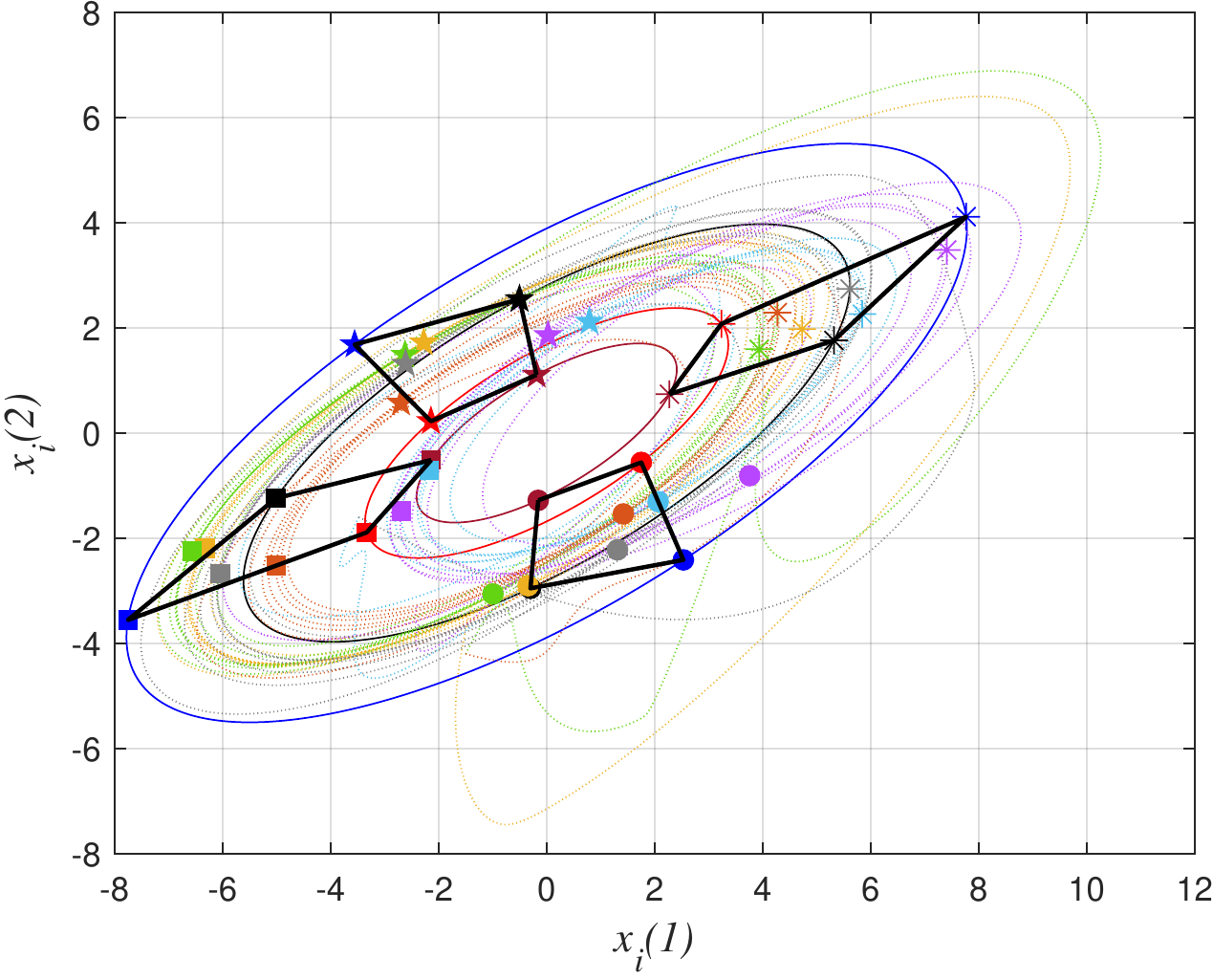}
\caption{Trajectories of the agents under the unbounded sensor and actuator attacks, using the proposed resilient control method.}
\label{Fig6}
\end{center} 
\end{figure}
\begin{figure}
\begin{center}
\includegraphics[scale=0.6]{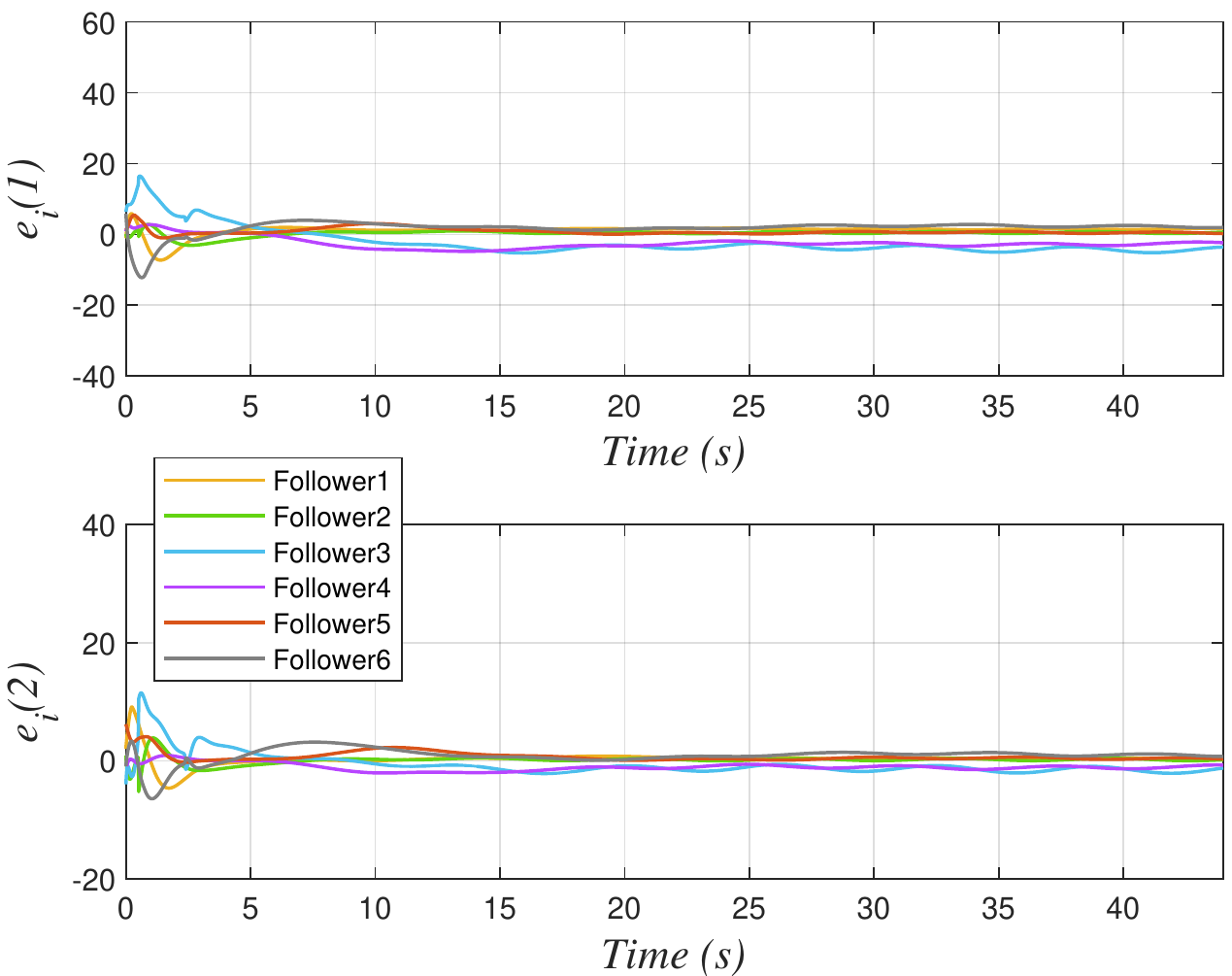}
\caption{Containment errors of the followers over time, using the proposed resilient control method.}
\label{Fig7}
\end{center} 
\end{figure} 

Next, we evaluate the system resilience against both unknown unbounded sensor and actuator attacks. Fig.~\ref{Fig4} shows the trajectories of the leaders and the followers under unbounded sensor and actuator attacks, using the conventional method. As seen, the trajectories of the followers diverge and are unbounded. From Lemma 2, the containment error in \eqref{eq7} characterizes the containment performance of the followers. Fig.~\ref{Fig5} illustrates the local containment error of each follower, using the conventional method. As seen, the local containment errors diverge and are unbounded. These validate that the conventional cooperative method fails to preserve the cooperative containment performance under unbounded sensor and actuator attacks. Fig.~\ref{Fig6} shows the trajectories of the agents using the proposed resilient method. As seen, the followers stay in a small neighborhood around the convex hull spanned by the leaders. Fig.~\ref{Fig7} shows the local containment error of each follower, using the proposed resilient method. As seen, the local containment errors are UUB. These reveal that the proposed resilient method successfully maintains the system stability and achieves the UUB containment convergence, even under simultaneously unknown unbounded sensor and actuator attacks.

\section{Conclusion}
In this paper, we have considered the resilient containment control problem of heterogeneous MAS against unbounded and correlated sensor attacks and generally unbounded actuator attacks. We have developed a distributed attack-resilient control framework to maintain the system stability and preserve the UUB convergence for the containment control performance. We have validated the efficacy of the proposed results using simulated case studies. Future work will study the more practical yet challenging scenario with generally unknown and unbounded sensor and actuator attacks, i.e., to relax Assumption 5.

\ifCLASSOPTIONcaptionsoff
  \newpage
\fi


\end{document}